\documentclass[12pt]{article}
\usepackage[utf8]{inputenc}
\usepackage[british]{babel}
\usepackage{cmap}
\usepackage{lmodern}
\usepackage[T1]{fontenc}

\usepackage{amssymb, amsmath, amsthm}
\usepackage[a4paper,top=25mm,bottom=25mm,left=25mm,right=25mm]{geometry}
\usepackage{ragged2e}
\usepackage{array, geometry, setspace, placeins}
\usepackage{authblk} 
\usepackage{pifont}
\usepackage{graphicx}
\usepackage[dvipsnames,svgnames,table]{xcolor}
\usepackage[figuresright]{rotating}
\usepackage{xtab} 
\usepackage{longtable} 
\usepackage{multirow}
\usepackage{footnote}
\usepackage[stable]{footmisc}
\usepackage{chngpage} 
\usepackage{pdflscape} 
\usepackage{tocbibind} 

\usepackage{pgfplots}
\pgfplotsset{compat=1.18}
\pgfplotsset{every tick label/.append style={font=\footnotesize}}
\usepackage{setspace}

\makesavenoteenv{tabular}
\usepackage{tabularx}
\usepackage{booktabs}
\usepackage{threeparttable} 
\usepackage[referable]{threeparttablex} 
\newcolumntype{R}{>{\raggedleft\arraybackslash}X}
\newcolumntype{L}{>{\raggedright\arraybackslash}X}
\newcolumntype{C}{>{\centering\arraybackslash}X}
\newcolumntype{A}{>{\columncolor{gray!25}}C}
\newcolumntype{a}{>{\columncolor{gray!25}}c}

\newlength{\tablen}

\usepackage{dcolumn} 
\newcolumntype{.}{D{.}{.}{-1}}

\usepackage{tikz}
\usetikzlibrary{arrows, calc, matrix, patterns, positioning, trees}
\usepackage[semicolon]{natbib}
\usepackage[hyphens]{url}
\usepackage{hyperref} 
\hypersetup{
  colorlinks   = true,    		
  urlcolor     = blue,    		
  linkcolor    = blue,			
  citecolor    = ForestGreen,		  	
}
\usepackage{microtype}
\usepackage[justification=centering]{caption} 

\usepackage[labelformat=simple]{subcaption}

\DeclareCaptionLabelFormat{parenthesis}{(#2)}
\makeatletter
\renewcommand\p@subfigure{\arabic{figure}.}
\makeatother

\DeclareCaptionLabelFormat{parenthesis}{(#2)}
\makeatletter
\renewcommand\p@subtable{\arabic{table}.}
\makeatother

\usepackage{alphalph}

\def\addlegendimage{\csname pgfplots@addlegendimage\endcsname}

\usepackage{enumitem}

\setlist[itemize]{leftmargin=2.5\parindent}
\setlist[enumerate]{leftmargin=2.5\parindent}

\theoremstyle{plain}

\theoremstyle{definition}

\theoremstyle{remark}

\def\keywords{\vspace{.5em} 
{\noindent \textit{Keywords}: }}

\def\AMS{\vspace{.5em} 
{\noindent \textbf{\emph{MSC} class}: }}

\def\JEL{\vspace{.5em} 
{\noindent \textbf{\emph{JEL} classification number}: }}

\title{Match forecasts in UEFA club competitions: \\ Elo ratings versus Transfermarkt valuations}
\author{
\href{https://sites.google.com/view/csurillag/}{Gergely Csurilla}\thanks{~E-mail: \emph{csurilla.gergely@krtk.elte.hu} \newline
Institute of Economics, ELTE Centre for Economic and Regional Studies, Budapest, Hungary \newline
Institute for Computer Science and Control (SZTAKI), Hungarian Research Network (HUN-REN), Laboratory on Engineering and Management Intelligence, Research Group of Operations Research and Decision Systems, Budapest, Hungary \newline
Hungarian University of Sports Science, Budapest, Hungary} $\qquad \qquad$
\href{https://sites.google.com/view/laszlocsato}{L\'aszl\'o Csat\'o}\thanks{~Corresponding author \newline
E-mail: \emph{laszlo.csato@sztaki.hun-ren.hu} \newline
Institute for Computer Science and Control (SZTAKI), Hungarian Research Network (HUN-REN), Laboratory on Engineering and Management Intelligence, Research Group of Operations Research and Decision Systems, Budapest, Hungary \newline 
Corvinus University of Budapest (BCE), Institute of Operations and Decision Sciences, Department of Operations Research and Actuarial Sciences, Budapest, Hungary}
}
\date{\today}

\def\Dedication{
{\noindent
``\emph{Second, it would be interesting to check whether the reported
results carry over to other settings in soccer. For example, can Transfermarkt values also generate accurate predictions for games between clubs?}'' \citep[p.~28]{Peeters2018}
}}

\begin{document}

\newgeometry{top=25mm,bottom=25mm,left=25mm,right=25mm}
\maketitle
\thispagestyle{empty}
\Dedication

\begin{abstract}
\noindent
The pre-season strengths of European football clubs are usually measured by two proxies in the literature. Football Club Elo Ratings provide strictly performance-based Elo ratings from the early days of the European Cups, while Transfermarkt valuations are crowd-based estimates of squad market values. This paper compares them by evaluating their ability to forecast the results of matches played in the UEFA Champions League and the UEFA Europa League between the seasons 2020/21 and 2024/25.
The two indicators yield almost identical out-of-sample accuracy when used separately. Combining the two measures leads to a modest improvement, but the best aggregation procedure is sensitive to the forecast target. Our results suggest that seeding based on Elo ratings would be (closely) optimal.
\end{abstract}


\AMS{62F07, 62P20, 90B90}

\JEL{C53, Z20}

\clearpage
\restoregeometry

\section{Introduction} \label{Sec1}

Simulating tournament designs strongly relies on measures of team strength \citep{LasekGagolewski2018, SziklaiBiroCsato2022, DevriesereCsatoGoossens2025}. Elo ratings \citep{Elo1978, Aldous2017, GomesdePinhoZancoSzczecinskiKuhnSeara2024} have become a natural benchmark as they summarise past results in a single dynamic, continuously updated value, and can be used across competitions and countries \citep{HvattumArntzen2010, vanEetveldeLey2019, Csato2024c, GerkenZhangGarnicaCaparrosGardewegMemmertWunderlich2026}.
The market value of the squad provides a different source of information. Transfermarkt values are crowd-based estimates of player market values and have been shown to contain meaningful information about player quality and transfer fees \citep{HermCallsen-BrackerKreis2014, CoatesParshakov2022}.

Crucially, \citet{Peeters2018} finds that Transfermarkt valuations outperform the previous FIFA World Ranking (used until 2018) and the World Football Elo Ratings in forecasting matches played by national football teams.
We ask whether the same market value signal is able to improve season-ahead predictions in the most prestigious European club football competitions---as suggested in the last sentence of \citet{Peeters2018}, quoted above.

This setting changes the forecasting problem in several ways. Club teams play significantly more games than national teams, and compete in different domestic leagues and cups in parallel to meeting in international competitions organised by the Union of European Football Associations (UEFA). The higher number of matches played potentially makes Elo ratings more informative, even though the set of matches forms a sparse, fragmented network \citep{GerkenZhangGarnicaCaparrosGardewegMemmertWunderlich2026}.
On the other hand, Elo ratings at the beginning of the season are insensitive to squad changes, as they depend only on past performance observed on the field. Since the squads of clubs change more frequently, Transfermarkt squad values before the season provide potentially valuable orthogonal information that might improve the accuracy of season-ahead forecasts. However, Transfermarkt values tend to underestimate transfer fees, and this bias is quite heterogeneous, which can reduce their predictive power \citep{CoatesParshakov2022}.

Besides focusing on club football, another crucial difference compared to \citet{Peeters2018} is that we deliberately use pre-season values: the Elo ratings observed on 1 September and the lagged Transfermarkt squad value from the previous season.
Our aim is not to construct the most accurate prediction model, but to explore how tournament organisers and researchers could predict the results of competitions based solely on the information available before they start. This also explains why betting odds are ignored despite their proven value in providing accurate forecasts \citep{WunderlichMemmert2018, Wunderlich2025}.

We aim to explore whether pre-season Football Club Elo Ratings and Transfermarkt values should be treated as substitutes or complements in predicting European club football results. We therefore compare Elo-only, Transfermarkt-only, and joint models, and consider combining model forecasts. This forecast-combination layer is important because a central lesson in forecasting is that simple averages are often difficult to beat \citep{BatesGranger1969}. Thus, our analysis also contributes to the question of whether pooling forecasts from strongly correlated signals based on different information sources could improve predictive accuracy when both signals are fixed at a common forecast origin.

The comparison covers realised goal difference, expected goals difference, and ordered match outcome. The expected goals target has been inspired by works that use shot quality to separate underlying performance from variation in finishing \citep{BrechotFlepp2020, MeadOHareMcMenemy2023}.
For probability forecasts, Rank Probability Score (RPS) is our headline scoring rule because it respects the natural ordering of win, draw, loss \citep{ConstantinouFenton2012, LeyvandeWieleVanEeetvelde2019}. Nevertheless, Brier scores, log loss, and classification success reported, too.

According to the results, both indicators contain predictive information in the common sample, and neither clearly dominates across all forecast targets. Combining their forecasts yields small improvements in aggregate accuracy. Estimated stacking records the lowest RMSE for the continuous targets, while equal-weight pooling implies the lowest RPS and log loss under both statistical models (ordered probit and multinomial logit). The ranking changes across evaluation seasons, and the probability-score advantage of pooling over Transfermarkt alone remains uncertain.

Our results could be useful especially for designing future competitions. Seeding, the ordering of teams prior to the tournament, plays a crucial rule in determining the outcome of any tournament except for round-robin leagues (where each team plays the same number of matches against any other team) \citep{ScarfYusof2011}. Misaligned seeding policies substantially bias both the FIFA Men's \citep{LaprePalazzolo2023} and Women's \citep{LaprePalazzolo2022} World Cups, as well as the UEFA European Championships \citep{LapreAmato2025}, by creating strongly imbalanced groups. Even though the new incomplete round-robin format of UEFA club competitions, introduced in the 2024/25 season, does not contain groups, seeding remains important as it determines the set of opponents, which has a substantial impact on the final ranking \citep{CsatoDevriesereGoossensGyimesiLambersSpieksma2026, CsatoGyimesiGoossensDevriesereLambersSpieksma2026, GuyonBenSalemBuchholtzerTanre2026}.


The remainder of the paper is organised as follows. Section~\ref{Sec2} gives an overview of
related studies. Section~\ref{Sec3} describes our data and forecast samples. Section~\ref{Sec4} presents the empirical design, Section~\ref{Sec5} reports the results, and Section~\ref{Sec6} concludes.

\section{Related literature} \label{Sec2}

Football forecasts are obtained either from score models or from direct models of the trichotomous win, draw, loss outcome. Poisson-type specifications provide the main score-based benchmark \citep{Maher1982, vanEetveldeLey2019}. \citet{Goddard2005} compares score and result models, while \citet{KoopmanLit2019} develop dynamic specifications for goals, goal difference, and the trichotomous categorical result.
Another class of models aggregates past performance in a team rating. \citet{HvattumArntzen2010} use Elo differences as covariates in ordered logit models, \citet{LeitnerZeileisHornik2010} embed ratings in tournament simulations, and \citet{LasekGagolewski2021} derive online rating updates from likelihood-based models. Related time-varying paired comparison methods are studied by \citet{BakerMcHale2015}, \citet{BakerMcHale2018}, and \citet{LeyvandeWieleVanEeetvelde2019}. Empirical evidence suggests that the leading score- and rating-based methods often have similar predictive accuracy \citep{LasekSzlavikBhulai2013, HubacekSourekZelezny2022}. We therefore use parsimonious models and focus on the information content of the strength measures, not on developing a new prediction method.

Team ratings based only on historical results incorporate the impact of player transfers only after they affect observed performance. Player-level models address this issue more directly. According to \citet{ArntzenHvattum2021}, Elo team ratings and plus-minus player ratings show similar performance separately, but a model containing both improves match forecasts. Analogously, \citet{HolmesMcHale2024} model team strength from the abilities of their players.

Transfermarkt squad values provide an aggregate measure of a related source of information, squad composition and assessed player quality. Crowd valuations predict realised transfer fees \citep{HermCallsen-BrackerKreis2014}, even though they suffer from systematic bias and substantial heterogeneity \citep{CoatesParshakov2022}. \citet{FranceschiBrocardFollertGouguet2024} survey the determinants of player valuation in football.

The current paper is also related to forecasting the results of UEFA club tournaments. \citet{CoronaForrestTenaHorrilloWiper2019} use a Bayesian match model, while \citet{DagaevRudyak2019}, \citet{Gyimesi2024}, and \citet{DevriesereGoossensSpieksma2026} use Poisson models to evaluate alternative Champions League formats and seeding rules. \citet{Csato2024c} compares the predictive power of UEFA club coefficients and Football Club Elo Ratings in the UEFA Champions League.

Combining forecasts is known to improve accuracy when the component models contain different information \citep{BatesGranger1969, WangHyndmanLiKang2023}. This motivates our distinction between a joint Elo--Transfermarkt model and pools of forecasts produced by separate models.

\section{Data and empirical setting} \label{Sec3}

The unit of observation is a match. The dataset covers the group stage, league phase, and knockout stage matches played in the UEFA Champions League and the UEFA Europa League in five recent seasons from 2020/21 to 2024/25. The group stage, used until the 2023/24 season, has been replaced by a new incomplete round-robin league phase in 2024/25.
Qualifying rounds are not considered. Three fixtures are excluded since they did not take place: Villarreal vs.\ Qaraba\u{g} in 2020/21, which UEFA awarded as a 3-0 win for Villarreal after being cancelled, and two RB Leipzig vs.\ Spartak Moscow fixtures in the 2021/22 Europa League Round of 16. The resulting panel contains 1,503 matches.

The match-level data come from FotMob (\url{https://www.fotmob.com/}), and contain match dates, home and away teams, final scores, and expected goals ($\mathit{xG}$). $\mathit{xG}$ is the likelihood that a shot is converted into a goal, based on the quality of the chance. We use these match records to construct three forecast targets.

Let $h$ and $a$ denote the home and away clubs, respectively, in match $i$ played in season $t$.
The realised goal difference is
\begin{equation}
\mathit{GD}_{it} = G^h_{it} - G^a_{it},
\end{equation}
where $G^h_{it}$ and $G^a_{it}$ are the number of goals scored by the home and away clubs, respectively.

Analogously, the expected goals difference is
\begin{equation}
\mathit{xGD}_{it} = xG^h_{it} - xG^a_{it}.
\end{equation}

The third target is the ordered match outcome:
\begin{equation}
R_{it}=\begin{cases}
0, & \text{if the away team wins}; \\
1, & \text{if the result is a draw}; \\
2, & \text{if the home team wins}.
\end{cases}
\end{equation}

The match-level $\mathit{xG}$ and final score on FotMob include potential extra time in the second leg of knockout matches. The baseline retains these observations, while a robustness sample excludes all matches where extra time is played.


The explanatory variables are two season-ahead measures of relative club strength. The first indicator is Football Club Elo Ratings (\url{http://clubelo.com/}), which provide Elo ratings for European football clubs and is widely used in the academic literature as a proxy of team strength \citep{BoskerGurtler2024, Csato2022b, Csato2024c, CsatoPetroczy2026, YildirimBilman2025a, YildirimBilman2025b}.
We use ratings on 1 September of the corresponding season for each club and define the home-away difference as:
\begin{equation}
\text{Elo}_{it} = \text{Elo}^h_t - \text{Elo}^a_t.
\end{equation}

The second indicator is the lagged Transfermarkt squad market value from the previous season. Transfermarkt (\url{https://www.transfermarkt.com/}) publishes football transfers, market values, and squad information; prior research interprets these values as crowd-informed assessments of player and squad quality \citep{HermCallsen-BrackerKreis2014, Peeters2018, CoatesParshakov2022}. 
For forecast season $t$, we use the club-level Transfermarkt value assigned to season $t-1$. This value is fixed across all matches of a club in season $t$.
The regressor is the log market-value difference:
\begin{equation}
\text{TM}_{it} = \log \left( V^h_{t-1} \right) - \log \left( V^a_{t-1} \right),
\end{equation}
where $V^h_{t-1}$ and $V^a_{t-1}$ denote the lagged squad values of the home and away clubs, respectively. Taking the difference in logarithms expresses relative squad strength, and limits the influence of clubs with exceptionally high market values.

For fixtures played on a neutral field, home and away correspond to the match record. We retain this orientation for every difference, but set the home venue indicator to zero.

The merge of data is performed in two steps. First, match-level FotMob records are standardised by season, competition, phase, date, home club and away club. Second, club-season indicator files are harmonised across naming conventions and joined separately to the home and away clubs. The final panel contains the realised outcome, $\mathit{xG}$ values, pre-season Elo ratings, and lagged Transfermarkt squad values for each fixture.

\begin{table}[t!]
\centering
\begin{threeparttable}
\caption{Coverage of pre-season strength indicators}
\label{Table1}

\begin{tabularx}{0.75\textwidth}{lCCc}
\toprule
Competition & \(N\) & Elo & Transfermarkt \\
\midrule
Champions League &       689 &     100\% &     98.3\% \\
Europa League &       814 &     100\% &     88.1\% \\
\bottomrule
\end{tabularx}
\begin{tablenotes}
\footnotesize
\item \textit{Notes:}
Entries except $N$ are sample shares. Cancelled and administratively awarded fixtures are excluded. Elo is observed on 1 September of the corresponding season; Transfermarkt is the lagged previous season squad value. Coverage requires values for both clubs.
\end{tablenotes}
\end{threeparttable}
\end{table}

The main sample requires a played match with non-missing outcome, $\mathit{xG}$, Elo, and Transfermarkt values. Table~\ref{Table1} reports coverage by competition. Outcome, $\mathit{xG}$, and Elo are observed for every match. Transfermarkt coverage is above 98\% in the Champions League and 88\% in the Europa League.
Since Elo coverage is complete, the Transfermarkt-covered matches form the common Elo--Transfermarkt sample of 1,394 observations. At the club-season level, 17 observations lack a lagged Transfermarkt value. Their mean Elo rating is 1487.3 and their maximum Elo rating is 1603.1 (Maccabi Haifa in 2022/23), compared with the mean of 1712.3 among the 336 covered club-season observations. Missing coverage is therefore concentrated among lower-rated clubs.

\begin{table}[t!]
\centering
\caption{Descriptive statistics, common Elo--Transfermarkt sample}
\label{Table2}

\begin{threeparttable}
\rowcolors{1}{}{gray!20}
\begin{tabularx}{\textwidth}{lRRRRR} \toprule
Variable & $N$ & Mean & Std.~dev. & Min & Max \\ \bottomrule
Goal difference &      1394 & \(    0.366\) & \(    1.970\) & \(   -6.000\) & \(    7.000\) \\
Expected goals difference &      1394 & \(    0.327\) & \(    1.387\) & \(   -3.820\) & \(    5.730\) \\
Elo difference &      1394 & \(   -0.090\) & \(  168.908\) & \( -613.805\) & \(  499.658\) \\
Log Transfermarkt difference &      1394 & \(    0.001\) & \(    1.353\) & \(   -4.078\) & \(    3.872\) \\ \bottomrule
\end{tabularx}
\begin{tablenotes}
\footnotesize
\item \textit{Notes:} Goal difference and expected goals difference are measured as home minus away. Elo difference is the 1 September home-away rating difference. Transfermarkt is the log difference of the lagged previous season squad values.
\end{tablenotes}
\end{threeparttable}
\end{table}

Table~\ref{Table2} summarises descriptive statistics for the sample. The positive means of home-away goal difference and home-away $\mathit{xG}$ difference indicate the well-documented home advantage \citep{Pollard2008}. The means of Elo and log Transfermarkt differences are essentially zero, since most of the matches are home-away, except for the finals and the league phase in the 2024/25 season.
The correlation between the Elo and the log Transfermarkt differences is 0.877, showing substantial but imperfect overlap between the two indicators.

Expanding-window forecasts are first generated for the 2022/23 season. This initial held-out forecast is used to estimate the data-driven combination parameters. The headline evaluation covers the 2023/24 and 2024/25 seasons, with 238 and 362 matches, respectively, giving 600 evaluation observations.

\begin{table}[t!]
\centering
\caption{Sample sizes under robustness restrictions}
\label{Table3}

\begin{threeparttable}
\rowcolors{1}{}{gray!20}
\begin{tabularx}{\textwidth}{lCCc} \toprule
Sample & Total $N$ & Evaluation $N$ & League-phase $N$ \\ \bottomrule
Main &      1394 &       600 &       272 \\
No extra time &      1364 &       585 &       272 \\
No finals &      1384 &       596 &       272 \\
No neutral matches &      1352 &       584 &       264 \\
No second leg &      1203 &       522 &       272 \\
No last matchday &      1238 &       538 &       238 \\
No second leg + no last matchday &      1047 &       460 &       238 \\
No 2024/25 league phase &      1122 &       328 &         0 \\
\bottomrule
\end{tabularx}
\begin{tablenotes}
\footnotesize
\item \textit{Notes:}
The main sample contains played matches with non-missing outcome, $\mathit{xG}$, Elo, and Transfermarkt variables. The final restriction removes only the 2024/25 league phase and retains the 2024/25 knockout stage.
\end{tablenotes}
\end{threeparttable}
\end{table}

Table~\ref{Table3} presents the sample sizes for headline evaluation and seven restricted samples. The first three address match duration and venue by excluding matches with extra time, finals, and all fixtures classified as neutral or relocated. The next three remove knockout second legs, last matchdays in the first phase of the competition, or both sets of these matches. The final restriction excludes the 2024/25 league phase, but retains the knockout stage from the same season.
Section~\ref{Sec44} will justify these restrictions.

The COVID-affected seasons require special treatment due to weaker home advantage \citep{BrysonDoltonReadeSchreyerSingleton2021, FischerHaucap2021}.
Venue and attendance information are used to construct two match-context controls, $H_{it}$ and $C_{it}$, respectively.
$H_{it} = 1$ if the designated home club plays at its usual home venue and zero otherwise. $H_{it} = 0$ for finals that are always played on a neutral field. Other potential relocations are identified when the recorded venue is at least 150 kilometres from the home venue of the club.
$C_{it}=1$ if an attendance restriction is identified for match $i$. All matches in the 2020/21 season are classified as restriction-affected because match-level attendance records are incomplete. From 2021/22 onward, recorded zero attendance and manually verified cases determine the classification; missing attendance alone does not trigger the indicator. Appendix~\ref{Sec_A2} describes how these variables enter the models and compares the match-level definition with the earlier season-wide classification.
The baseline restriction contains 308 matches in the common sample: all (300) matches from the 2020/21 season and eight matches from the 2021/22 season. The interaction $H_{it}C_{it}$ equals one for 293 matches.

\section{Methodology} \label{Sec4}

The methodology of our empirical analysis is presented in the following structure: model specification (Section~\ref{Sec41}), out-of-sample forecast design and evaluation (Section~\ref{Sec42}), forecast combination (Section~\ref{Sec43}), and sensitivity analysis by reasonable sample restrictions (Section~\ref{Sec44}).

\subsection{Baseline outcome models} \label{Sec41}

For a continuous outcome $Y_{it} \in \{\mathit{GD}_{it}, \mathit{xGD}_{it}\}$, the linear in-sample specification, estimated by ordinary least squares (OLS), is
\begin{equation}
Y_{it} = \alpha + X_{it}'\beta + \gamma_H H_{it} + \gamma_C H_{it}C_{it} + \delta_{c(i)} + \rho_{p(i)} + \lambda_t + u_{it}.
\label{eq:ols}
\end{equation}
$X_{it}$ contains Elo, Transfermarkt, or both strength indicators. The coefficient $\gamma_H$ measures the home venue effect in matches without an identified attendance restriction, $\gamma_H + \gamma_C$ is the corresponding effect under restrictions. Competition, phase, and season fixed effects are denoted by $\delta_{c(i)}$, $\rho_{p(i)}$, and $\lambda_t$, respectively. Phase fixed effects distinguish the group or league phase from the knockout stage. The in-sample specifications use standardised strength indicators to facilitate coefficient comparisons.
Standard errors are clustered by unordered team-pair-season.

For the ordered match result, the baseline probability model is an ordered probit. The latent home team performance index is
\begin{equation}
R^\ast_{it} = X_{it}'\beta + \gamma_H H_{it} + \gamma_C H_{it}C_{it} +\delta_{c(i)} + \rho_{p(i)} + \lambda_t + \varepsilon_{it},
\qquad
\varepsilon_{it} \sim N(0,1).
\label{eq:oprobit-latent}
\end{equation}
The observed result is generated by two estimated cutpoints:
\begin{equation}
R_{it}=j \quad \text{if} \quad \kappa_{j-1}<R^*_{it}\leq \kappa_j, \qquad j=0,1,2,
\end{equation}
with $\kappa_{-1}=-\infty$ and $\kappa_2=\infty$. 
We estimate the two cutpoints freely. Ordered probit is used because the three outcomes correspond to negative, zero, and positive goal differences from the perspective of the designated home team. The category codes do not impose equal spacing between outcomes. No separate intercept is included because the location of the latent index is absorbed by the estimated cutpoints. Although raw Elo difference enters the latent index linearly, the ordered probit link maps that index nonlinearly into the three outcome probabilities.

The ordered probit model uses a common coefficient vector for the two cumulative probabilities, which bounds how the predictors affect the probability of draw. To check the robustness of the forecasting conclusions, we also estimate multinomial logit models, which do not impose this common-coefficient restriction. The comparison uses the same predictors and the out-of-sample forecast design (see Section~\ref{Sec42} and Table~\ref{Table5}).

\subsection{Forecast design and evaluation} \label{Sec42}

Forecasts are generated recursively; for each forecast season $t$, model parameters are estimated from all seasons $\tau < t$. The Elo and Transfermarkt measures, together with the estimated coefficients, are fixed throughout season $t$. Forecasts are conditional on the fixtures and match contexts that subsequently occur. Competition, phase, venue, and attendance-regime variables describe the match being forecast, but no outcome, $\mathit{xG}$, or within-season performance information from season $t$ is used to re-estimate or update the models. Season fixed effects are omitted because the effect for the forecast season is unavailable at the forecast origin.
Forecasts for the 2022/23 season provide the initial held-out observations used to estimate the data-driven combination parameters; the headline evaluation begins in 2023/24.

We use a deliberately simple historical benchmark as a na\"ive reference forecast.
For continuous outcomes, it is the historical mean with an expanding window:
\begin{equation}
\hat{Y}^{\mathit{hist}}_{it} = \frac{1}{|\mathcal{I}_t|}\sum_{\ell\in\mathcal{I}_t}Y_{\ell},
\end{equation}
where $\mathcal{I}_t$ is the set of training matches available before season $t$.
For the ordered match outcome, it uses expanding-window outcome frequencies with Laplace smoothing:
\begin{equation}
\hat{p}^{\mathit{hist}}_{jt}=\frac{n_{jt}+1}{n_t+3}, \qquad j=0,1,2,
\end{equation}
where $n_{jt}$ is the number of training matches with outcome $j$.
Since it does not include the match-context controls of the main strength models, the incremental comparison between Elo and Transfermarkt is based primarily on specifications estimated with the same set of controls.

For the trichotomous ordered match outcome, the ordered-probit forecasts are complemented by multinomial logit forecasts. With draw as the base outcome, the multinomial logit probabilities are
\begin{equation}
P(R_{it} = j\mid X_{it})=\frac{\exp(\eta_{jit})}{\sum_{k=0}^2 \exp(\eta_{kit})}, \qquad j=0,1,2,
\end{equation}
where $\eta_{1it}=0$ for the base outcome. This robustness check avoids imposing the single-index structure of the ordinal model on the draw probability.

Continuous forecasts are evaluated by the root mean squared error (RMSE) and the out-of-sample $R^2$ relative to the historical benchmark:
\begin{equation}
R^2_{OOS}=1-\frac{\sum_{i\in \mathcal{T}}(Y_i-\hat{Y}_i)^2}{\sum_{i\in \mathcal{T}}(Y_i-\hat{Y}^{\mathit{hist}}_i)^2},
\label{eq:oosr2}
\end{equation}
where $\mathcal{T}$ is the forecast-evaluation sample.

Probability forecasts are evaluated with proper scoring rules \citep{GneitingRaftery2007}. Our primary measure is the ranked probability score \citep{Epstein1969}:
\begin{equation}
\mathit{RPS} = \frac{1}{2N}\sum_{i=1}^N\sum_{k=0}^1\left(\sum_{j=0}^{k}\hat{p}_{ij}-\mathbf{1}\{R_i\leq k\}\right)^2.
\label{eq:rps}
\end{equation}
RPS evaluates cumulative probabilities across away wins, draws, and home wins \citep{ConstantinouFenton2012, LeyvandeWieleVanEeetvelde2019}. Lower values indicate better forecasts. 
In the following, all tables and the accompanying discussion report its value multiplied by 100. Pairwise RPS loss differentials and their standard errors are presented on the same scale. Naturally, this rescaling has no effect on model rankings or statistical tests.

Alongside RPS, we also report the Brier score \citep{Brier1950}:
\begin{equation}
\mathit{BS} = \frac{1}{3N}\sum_{i=1}^N\sum_{j=0}^2\left(\hat{p}_{ij}-\mathbf{1}\{R_i=j\}\right)^2;
\label{eq:brier}
\end{equation}
and the log loss:
\begin{equation}
\mathit{LL} = -\frac{1}{N}\sum_{i=1}^N\log(\hat{p}_{iR_i}).
\label{eq:logloss}
\end{equation}
Finally, the success rate is the proportion of matches in which the most likely outcome is realised. It is not a proper scoring rule and is interpreted as descriptive only.

Pairwise forecast comparisons are based on match-level loss differentials. For models $A$ and $B$, we define 
\begin{equation}
\Delta_i=L_{i,A}-L_{i,B}, 
\end{equation} 
so that negative values favour model $A$. The mean loss differential is estimated with an intercept-only regression. Each cluster groups matches between the same unordered team pair within a season and competition. Our two-sided $t$-tests and 95\% confidence intervals use the cluster-based standard errors and degrees of freedom. This follows the loss-differential framework of \citet{DieboldMariano1995} with a variance estimator for the match panel. The intervals describe uncertainty in mean loss differences for the fitted forecasts; they do not incorporate re-estimation of the outcome models or pooling weights. For the continuous targets, the comparisons use squared error losses, not RMSE differences. We also report forecast scores separately for the two evaluation seasons.

Calibration is assessed for the probability of home win. We cluster the predicted probabilities into bins and compare the mean predicted probability with the observed frequency in each bin. Positive deviations of the observed frequency from the predicted probability indicate underestimation of home wins in the given probability bin.

\subsection{Forecast pooling} \label{Sec43}

Joint estimation is distinguished from forecast pooling. The joint Elo--Transfermarkt model includes both strength indicators in the same outcome equation. Forecast pools estimate the Elo-only and Transfermarkt-only models separately and combine their predictions.

For a continuous outcome, the equal-weight pool is
\begin{equation}
\hat{Y}^{eq}_{it} = 0.5 \cdot \hat{Y}^{\mathit{Elo}}_{it} + 0.5 \cdot \hat{Y}^{\mathit{TM}}_{it}.
\label{eq:avg-cont}
\end{equation}
For the outcome probabilities, pooling is applied outcome by outcome:
\begin{equation}
\hat{p}^{eq}_{ijt} = 0.5 \cdot \hat{p}^{\mathit{Elo}}_{ijt} + 0.5 \cdot \hat{p}^{\mathit{TM}}_{ijt},
\qquad j=0,1,2.
\label{eq:avg-prob}
\end{equation}
Equal weighting is specified in advance and does not require any estimation.

We also construct data-driven combinations from earlier out-of-sample forecasts. For continuous outcomes, the stacking specification is
\begin{equation}
\hat{Y}^{stack}_{it} = a_t + b_{E,t}\hat{Y}^{\mathit{Elo}}_{it} + b_{T,t}\hat{Y}^{\mathit{TM}}_{it}.
\end{equation}
The intercept and slope coefficients are unconstrained; the two slopes need not be positive or sum to one. For the outcome probabilities, we select
$\omega_t\in\{0,0.05,\ldots,1\}$ to minimise RPS over earlier
out-of-sample forecasts:
\begin{equation}
\hat{p}^{rps}_{ijt} = (1-\omega_t) \cdot \hat{p}^{\mathit{Elo}}_{ijt} + \omega_t \cdot \hat{p}^{\mathit{TM}}_{ijt}.
\end{equation}
All estimated parameters are based exclusively on forecasts and outcomes available before the current forecast season.

For the 2023/24 forecasts, we estimate combination parameters from the 238 out-of-sample forecasts for 2022/23. For 2024/25, this sample expands to 476 forecasts from 2022/23 and 2023/24.
Appendix Table~\ref{Table_A1} reports the estimated parameters, including the stacking intercepts.

\subsection{Robustness checks} \label{Sec44}

Our sensitivity analyses examine four features of the underlying data that may affect forecast performance: match duration and venue, strategic match context, the replacement of the group stage with the league phase, and selected modelling and measurement choices. 

First, the FotMob final score and match-level $\mathit{xG}$ include extra time if it is played. Hence, we repeat the forecast evaluation after excluding all matches with extra time. Venue may also affect the interpretation of the home indicator. Finals remain in the baseline sample but are treated as neutral ($H_{it}=0$). We report separate results after excluding finals and after removing all fixtures classified as neutral or relocated (see Section~\ref{Sec3}).

Second, the incentives in some matches may be distorted by strategic considerations. The second leg of a knockout tie is a prominent example: the relevant objective of any club is advancement over the two matches, not the result of the second leg in isolation. Therefore, a draw or even a small loss may be sufficient if it leads to qualification.
The last matchday in the group stage and the league phase could involve weak or asymmetric incentives once qualification or elimination has already been determined \citep{CsatoMolontayPinter2024, DevriesereGoossensSpieksma2026, Gyimesi2024}; even deliberate losing may arise under particular conditions \citep{Rewilak2026}. We consequently report RPS results after excluding knockout second legs, after excluding the last matchday in the first phase, and after applying both restrictions jointly.

Third, the 2024/25 season introduced an incomplete round-robin league phase instead of the previous group stage, which might affect the probability of a draw \citep{Csato2025h, WinkelmannDeutscherMichels2026}.
Thus, one forecast exercise excludes the 2024/25 league-phase matches but retains the knockout stage from the same season.
In a separate analysis, we add the 2024/25 league phase indicator and its interactions with both strength indicators to the joint in-sample models. The baseline controls and fixed effects are retained, and the two interaction coefficients are tested jointly and separately.

Finally, two supplementary specifications are investigated in the Appendix.
The raw Elo difference is compared with the canonical Elo expected score, both without a home adjustment and with a forecast-origin-specific adjustment estimated exclusively from earlier unrestricted-attendance matches. This might be important since the update of the Elo rating is not a linear function of Elo difference \citep{CsatoPetroczy2025a, CsatoPetroczy2026}.
Second, as restrictions on the number of spectators influenced home advantage in European football \citep{BrysonDoltonReadeSchreyerSingleton2021, FischerHaucap2021}, we compare the match-level crowd-restriction indicator with the earlier season-wide classification.


\section{Results} \label{Sec5}

Section~\ref{Sec51} examines the in-sample associations between the strength indicators and match outcomes. Section~\ref{Sec52} compares out-of-sample accuracy, tests the significance of loss differences, and assesses sensitivity to sample composition and the league-phase format.
Calibration of home wins is addressed in Appendix~\ref{Sec_A1}, and additional specification tests are conducted in Appendix~\ref{Sec_A2}.

\subsection{In-sample baseline estimates} \label{Sec51}

\begin{table}[t!]
\centering
\caption{In-sample explanatory models using Elo and Transfermarkt}
\label{Table4}

\begin{subtable}{\textwidth}
\centering
\caption{Goal difference}
\label{Table4a}
\begin{tabularx}{0.8\textwidth}{lCCC}
\toprule
 & Elo & Transfermarkt & Joint Elo--TM \\
\midrule
\multirow{2}{*}{Elo} & 0.977*** & \multirow{2}{*}{---} & 0.662*** \\
 & (0.047) &  & (0.099) \\
\multirow{2}{*}{Transfermarkt} & \multirow{2}{*}{---} & 0.938*** & 0.358*** \\
 &  & (0.047) & (0.095) \\
\midrule
\(N\) &      1394 &      1394 &      1394 \\
Adjusted \(R^2\) &     0.254 &     0.236 &     0.261 \\
\bottomrule
\end{tabularx}
\end{subtable}

\vspace{0.5cm}
\begin{subtable}{\textwidth}
\centering
\caption{Expected goals difference}
\label{Table4b}
\begin{tabularx}{0.8\textwidth}{lCCC}
\toprule
 & Elo & Transfermarkt & Joint Elo--TM \\
\midrule
\multirow{2}{*}{Elo} & 0.746*** & \multirow{2}{*}{---} & 0.409*** \\
 & (0.032) &  & (0.066) \\
\multirow{2}{*}{Transfermarkt} & \multirow{2}{*}{---} & 0.742*** & 0.384*** \\
 &  & (0.032) & (0.063) \\
\midrule
\(N\) &      1394 &      1394 &      1394 \\
Adjusted \(R^2\) &     0.297 &     0.295 &     0.314 \\
\bottomrule
\end{tabularx}
\end{subtable}

\vspace{0.5cm}
\begin{subtable}{\textwidth}
\centering
\caption{Ordered match result}
\label{Table4c}
\begin{threeparttable}
\begin{tabularx}{0.8\textwidth}{lCCC}
\toprule
 & Elo & Transfermarkt & Joint Elo--TM \\
\midrule
\multirow{2}{*}{Elo} & 0.596*** & \multirow{2}{*}{---} & 0.402*** \\
 & (0.036) &  & (0.070) \\
\multirow{2}{*}{Transfermarkt} & \multirow{2}{*}{---} & 0.568*** & 0.223*** \\
 &  & (0.037) & (0.071) \\
\midrule
\(N\) &      1394 &      1394 &      1394 \\
Pseudo \(R^2\) &     0.108 &     0.100 &     0.112 \\
\bottomrule
\end{tabularx}
\begin{tablenotes}
\footnotesize
\item
\textit{Notes:} Coefficients are based on standardised season-fixed strength indicators. All specifications include the home venue indicator and its interaction with the baseline crowd-restriction indicator. All three outcome models include season, competition, and phase fixed effects and use standard errors clustered by unordered team-pair-season within competition. The ordered match result is estimated by ordered probit.  Standard errors are in parentheses. Cells with -- indicate that the indicator is not included in the specification.
Significance: * $p<0.10$, ** $p<0.05$, *** $p<0.01$.
\end{tablenotes}
\end{threeparttable}
\end{subtable}
\end{table}


The in-sample estimates for the common Elo--Transfermarkt sample are reported in Table~\ref{Table4}.
Table~\ref{Table4a} shows that both standardised indicators are positively associated with realised goal difference, and both estimates are statistically significant. The single-indicator coefficients are similar in magnitude, 0.977 for Elo and 0.938 for Transfermarkt. When both indicators are included, the coefficients decline to 0.662 and 0.358, respectively, but remain statistically significant. The adjusted $R^2$ increases from 0.254 in the Elo-only model and 0.236 in the Transfermarkt-only model to 0.261 in the joint specification. The two measures contain substantial common information about realised score differences, with a limited additional contribution from their joint use.

The expected goals difference results in Table~\ref{Table4b} provide stronger evidence of distinct conditional information. The Elo-only and Transfermarkt-only coefficients are nearly identical, at 0.746 and 0.742, and the two models have similar adjusted $R^2$ values. In the joint specification, the standardised coefficients remain comparable (0.409 for Elo and 0.384 for Transfermarkt), and both are significant. The adjusted $R^2$ rises from 0.297 and 0.295 in the single-indicator models to 0.314 in the joint model. Thus, each indicator contributes to explaining differences in chance creation after conditioning on the other. The increase in model fit is larger than for realised goal difference, although it remains moderate.

Table~\ref{Table4c} shows a similar pattern for ordered match results. Both Elo and Transfermarkt shift the latent index towards a more favourable outcome for the home club, and both coefficients remain statistically significant in the joint ordered-probit model. The conditional Elo coefficient is 0.402, compared with 0.223 for Transfermarkt. The pseudo-$R^2$, however, increases only marginally in the joint model.

To conclude, the in-sample results therefore indicate that neither indicator is conditionally redundant, but the gains in model fit are quite limited.

\subsection{Out-of-sample forecasts} \label{Sec52}

The out-of-sample analysis in Section~\ref{Sec521} uncovers whether the conditional associations in Table~\ref{Table4} translate into forecast improvements. The statistical differences between the prediction models are tested in Section~\ref{Sec522}.
Section~\ref{Sec523} presents the sensitivity analyses with various sample restrictions, and Section~\ref{Sec524} investigates the effect of the novel incomplete round-robin league phase on the estimated coefficients.

\subsubsection{Forecast accuracy} \label{Sec521}

\begin{table}[t!]
\centering
\caption{Out-of-sample forecast performance, common Elo--Transfermarkt sample}
\label{Table5}

\begin{subtable}{\textwidth}
\centering
\caption{Goal difference forecasts}
\label{Table5a}
\rowcolors{1}{}{gray!20}
\begin{tabularx}{0.6\textwidth}{lCCC} \toprule
Model & $N$ & RMSE & $R^2$ \\ \bottomrule
Historical &       600 &     1.958 &     0.000 \\
Elo &       600 &     1.710 &     0.237 \\
Transfermarkt &       600 &     1.710 &     0.237 \\
Joint Elo--TM &       600 &     1.698 &     0.248 \\
Equal-weight pool &       600 &     1.695 &     0.251 \\
Estimated combination &       600 &     1.686 &     0.259 \\
\bottomrule
\end{tabularx}
\end{subtable}

\vspace{0.5cm}
\begin{subtable}{\textwidth}
\centering
\caption{Expected goals difference forecasts}
\label{Table5b}
\rowcolors{1}{}{gray!20}
\begin{tabularx}{0.6\textwidth}{lCCC} \toprule
Model & $N$ & RMSE & $R^2$ \\ \bottomrule
Historical &       600 &     1.386 &     0.000 \\
Elo &       600 &     1.183 &     0.272 \\
Transfermarkt &       600 &     1.186 &     0.268 \\
Joint Elo--TM &       600 &     1.173 &     0.284 \\
Equal-weight pool &       600 &     1.172 &     0.285 \\
Estimated combination &       600 &     1.168 &     0.290 \\
\bottomrule
\end{tabularx}
\end{subtable}

\vspace{0.5cm}
\begin{subtable}{\textwidth}
\centering
\caption{Ordered match result forecasts, ordered probit}
\label{Table5c}
\rowcolors{1}{}{gray!20}
\begin{tabularx}{\textwidth}{lCCCCC} \toprule
Model & $N$ & $100 \times$RPS & Brier & Log loss & Success \\ \bottomrule
Historical &       600 &    23.068 &     0.207 &     1.029 &     0.510 \\
Elo &       600 &    19.729 &     0.183 &     0.932 &     0.558 \\
Transfermarkt &       600 &    19.533 &     0.182 &     0.928 &     0.577 \\
Joint Elo--TM &       600 &    19.539 &     0.182 &     0.926 &     0.560 \\
Equal-weight pool &       600 &    19.431 &     0.182 &     0.923 &     0.568 \\
Estimated combination &       600 &    19.480 &     0.182 &     0.925 &     0.563 \\ \bottomrule
\end{tabularx}
\end{subtable}

\vspace{0.5cm}
\begin{subtable}{\textwidth}
\centering
\caption{Ordered match result forecasts, multinomial logit}
\label{Table5d}
\begin{threeparttable}
\rowcolors{1}{}{gray!20}
\begin{tabularx}{\textwidth}{lCCCCC} \toprule
Model & \(N\) & $100 \times$RPS & Brier & Log loss & Success \\ \bottomrule
Historical &       600 &    23.068 &     0.207 &     1.029 &     0.510 \\
Elo &       600 &    19.761 &     0.184 &     0.935 &     0.565 \\
Transfermarkt &       600 &    19.524 &     0.182 &     0.928 &     0.575 \\
Joint Elo--TM &       600 &    19.579 &     0.183 &     0.929 &     0.562 \\
Equal-weight pool &       600 &    19.452 &     0.182 &     0.925 &     0.570 \\
Estimated combination &       600 &    19.505 &     0.182 &     0.927 &     0.563 \\ \bottomrule
\end{tabularx}
\begin{tablenotes}
\footnotesize
\item
\textit{Notes:}
Forecasts use an expanding window and are not updated within a forecast season. Lower RMSE, RPS, Brier score, and log loss are better, whereas higher $R^2$ and success are better. Out-of-sample $R^2$ is measured relative to the expanding-window historical benchmark. All models use the same training and evaluation matches at each forecast origin. Success is the proportion of matches in which the most likely outcome occurs. Exact ties follow the code order: home win, draw, away win.
\end{tablenotes}
\end{threeparttable}
\end{subtable}
\end{table}

According to Table~\ref{Table5}, all of our strength models clearly outperform the unconditional historical benchmark. Since the benchmark does not contain match-context controls, the substantive comparison concerns the Elo, Transfermarkt, joint, and pooled specifications, which use a common set of controls.

For goal difference (Table~\ref{Table5a}), the Elo-only and Transfermarkt-only models have exactly the same RMSE at the reported precision. Both joint estimation and forecast pooling reduce the error. The estimated combination records the lowest RMSE, followed by the equal-weight pool and the joint model. Their out-of-sample $R^2$ values are around 0.25. Thus, combining the two information sources improves goal difference forecasts, although the gain is small.

The expected goals difference results (Table~\ref{Table5b}) follow the same pattern. Elo performs slightly better than Transfermarkt when they are used separately. The joint model and both forecast-level combinations improve on the two single-indicator specifications. The estimated combination again records the lowest RMSE (1.168) and out-of-sample $R^2$ (0.290). The equal-weight pool is close, with an RMSE of 1.172 and an $R^2$ of 0.285. The combination gain also appears for the balance of chance creation, but estimating the stacking parameters yields only a minor numerical improvement over equal weighting.

For the outcome probabilities (Tables~\ref{Table5c} and \ref{Table5d}), the equal-weight pool has the lowest RPS under both probability models, but all RPS values are close to each other except for the na\"ive approach.
Log loss also favours equal weighting, while Brier score differences are negligible among the leading models. Classification success is the best for the Transfermarkt-only model under both links, illustrating that modal-outcome accuracy and probability-forecast quality do not always produce the same ranking. The estimated probability pools assign weights of 0.35--0.40 to Transfermarkt across the two forecast seasons (Table~\ref{Table_A1}), and have slightly worse aggregate RPS than the equal-weight pools.

Across the full evaluation sample, estimated stacking records the lowest RMSE for both continuous targets, while equal-weight pooling records the lowest RPS and log loss under both probability models. These are rankings of observed losses, not evidence that one combination method dominates all alternatives.
Table~\ref{Table_A2} in the Appendix shows that the model ranking changes across evaluation seasons. The joint model records the lowest RPS in the 2023/24 season, while Transfermarkt alone has the lowest RPS in the 2024/25 season under both probability specifications. Equal-weight pooling shows the best RPS over the full 600-match evaluation sample, but not in either season separately. The advantage of estimated stacking for the continuous targets also varies across seasons.


\subsubsection{Forecast comparison tests} \label{Sec522}

\begin{table}[t!]
\centering
\caption{Pairwise comparison tests of ordered probit forecasts based on RPS, main sample}
\label{Table6}

\centerline{
\begin{threeparttable}
\rowcolors{1}{}{gray!20}
\begin{tabularx}{1.1\textwidth}{lCCcC} \toprule
Comparison & $100\times$RPS $\Delta$ & $100\times$Std.e. & 95\% CI & $p$-value \\ \bottomrule \showrowcolors
Equal-weight vs.\ Elo & \(-0.297\) &     0.148 & \([-0.588,\,-0.007]\) & 0.045** \\
Equal-weight vs.\ TM & \(-0.102\) &     0.150 & \([-0.397,\,+0.194]\) & 0.499 \\
Estimated vs.\ Equal-weight & \(+0.049\) &     0.041 & \([-0.033,\,+0.130]\) & 0.240 \\
Estimated vs.\ Joint Elo--TM & \(-0.059\) &     0.048 & \([-0.153,\,+0.035]\) & 0.220 \\
Joint Elo--TM vs.\ Elo & \(-0.190\) &     0.075 & \([-0.337,\,-0.042]\) & 0.012** \\
Joint Elo--TM vs.\ Equal-weight & \(+0.108\) &     0.083 & \([-0.056,\,+0.272]\) & 0.198 \\
Joint Elo--TM vs.\ TM & \(+0.006\) &     0.230 & \([-0.446,\,+0.458]\) & 0.980 \\
TM vs.\ Elo & \(-0.195\) &     0.297 & \([-0.778,\,+0.387]\) & 0.510 \\ \bottomrule
\end{tabularx}
\begin{tablenotes} \footnotesize
\item \textit{Notes:}
The number of observations is 600 in all cases.
RPS $\Delta$ is the score of model A minus the score of model B; negative values favour model A. Differences, standard errors and confidence intervals report RPS multiplied by 100. Standard errors are clustered by unordered team-pair-season within competition. The 95\% confidence intervals use the same $t$ distribution and degrees of freedom as the reported tests. Equal-weight and Estimated denote uniform and estimated forecast pools. The intervals condition on the fitted forecasts; they do not re-estimate models or pooling weights.
Significance: * $p<0.10$, ** $p<0.05$, *** $p<0.01$.
\end{tablenotes}
\end{threeparttable}
}
\end{table}

According to Table~\ref{Table6}, both the equal-weight pool and the joint model achieve modest but statistically significant RPS gains over Elo. On the other hand, neither improves significantly on Transfermarkt, and the tests do not reject equal predictive accuracy for the two single-indicator models. Estimating the pool weights does not provide a significant improvement over equal weighting. 

The comparison with Transfermarkt differs for expected goals forecasts, as equal-weight pooling significantly reduces RMSE loss (Table~\ref{Table_A3}). Estimated stacking does not significantly improve on equal weighting for either continuous target. The evidence for pooling therefore depends on the forecast target and the loss function.

\subsubsection{Robustness to sample composition} \label{Sec523}

\begin{table}[t!]
\centering
\caption{Robustness checks for outcome probability forecasts}
\label{Table7}

\begin{subtable}{\textwidth}
\centering
\caption{Match duration and venue restrictions}
\label{Table7a}
\rowcolors{1}{}{gray!20}
\begin{tabularx}{\textwidth}{lCcCC} \toprule \showrowcolors
Model & Main & No extra time & No finals & No neutral \\ \bottomrule
Historical &    23.068 &    23.113 &    23.057 &    22.951 \\
Elo &    19.729 &    19.698 &    19.679 &    19.640 \\
Transfermarkt &    19.533 &    19.462 &    19.503 &    19.462 \\
Joint Elo--TM &    19.539 &    19.495 &    19.497 &    19.467 \\
Equal-weight pool &    19.431 &    19.380 &    19.391 &    19.353 \\
Estimated combination &    19.480 &    19.437 &    19.417 &    19.400 \\ \midrule \hiderowcolors
Observations &       600 &       585 &       596 &       584 \\ \bottomrule
\end{tabularx}
\end{subtable}

\vspace{0.5cm}
\begin{subtable}{\textwidth}
\centering
\caption{Strategic-context and format restrictions}
\label{Table7b}
\begin{threeparttable}
\rowcolors{1}{}{gray!20}
\begin{tabularx}{\textwidth}{lCCCCC} \toprule \showrowcolors
Model & Main & No second leg & No last matchday & Both & No league phase \\ \bottomrule
Historical &    23.068 &    23.287 &    23.131 &    23.365 &    23.135 \\
Elo &    19.729 &    19.819 &    19.839 &    19.956 &    19.410 \\
Transfermarkt &    19.533 &    19.646 &    19.644 &    19.788 &    19.232 \\
Joint Elo--TM &    19.539 &    19.623 &    19.636 &    19.750 &    19.205 \\
Equal-weight pool &    19.431 &    19.524 &    19.536 &    19.654 &    19.154 \\
Estimated combination &    19.480 &    19.560 &    19.546 &    19.675 &    19.198 \\ \midrule \hiderowcolors
Observations &       600 &       522 &       538 &       460 &       328 \\
\bottomrule
\end{tabularx}
\begin{tablenotes} \footnotesize
\item
\textit{Notes:}
Entries are the RPSs of ordered probit multiplied by 100; lower values are better. Finals are retained in the main sample but coded as neutral. The neutral or relocated restriction removes matches at venues at least 150 km from the venue of the home club. Column Both in panel (b) excludes knockout second legs and last first phase matchday in the first phase. The last column in panel (b) removes the 2024/25 league phase, but not the knockout stage. Each restriction applies to both training and evaluation matches. We re-estimate the component models and combination parameters at each forecast origin.
\end{tablenotes}
\end{threeparttable}
\end{subtable}
\end{table}

Table~\ref{Table7} shows that the ranking of prediction models with respect to their RPSs is stable across all sample restrictions considered. The equal-weight pool performs the best after excluding matches with extra time, finals, and all neutral or relocated fixtures (Table~\ref{Table7a}).

The strategic-context restrictions lead to the same ranking again (Table~\ref{Table7b}). The equal-weight pool continues to have the lowest RPS after excluding knockout second legs, the last matchday in the first phase, or both sets of matches. The last restriction removes the 2024/25 league phase. In the resulting evaluation sample of 328 matches, the equal-weight pool again records the best RPS. Although the level of forecast loss varies with sample composition, the comparison among the leading models remains stable.

\subsubsection{League-phase slope diagnostics} \label{Sec524}

\begin{table}[t!]
\centering
\caption{Structural break diagnostics for the 2024/25 league phase}
\label{Table8}

\begin{subtable}{\textwidth}
\centering
\caption{Goal difference}
\label{Table8a}
\rowcolors{1}{}{gray!20}
\begin{tabularx}{0.6\textwidth}{llC} \toprule
Estimator & Test & $p$-value \\ \bottomrule \showrowcolors
OLS & Joint Elo--TM slopes & 0.709 \\
OLS & Elo slope & 0.866 \\
OLS & Transfermarkt slope & 0.591 \\ \toprule
\end{tabularx}
\end{subtable}

\vspace{0.5cm}
\begin{subtable}{\textwidth}
\centering
\caption{Expected goals difference}
\label{Table8b}
\rowcolors{1}{}{gray!20}
\begin{tabularx}{0.6\textwidth}{llC} \toprule
Estimator & Test & $p$-value \\ \bottomrule \showrowcolors
OLS & Joint Elo--TM slopes & 0.496 \\
OLS & Elo slope & 0.238 \\
OLS & Transfermarkt slope & 0.323 \\ \toprule
\end{tabularx}
\end{subtable}

\vspace{0.5cm}
\begin{subtable}{\textwidth}
\centering
\caption{Ordered match result}
\label{Table8c}
\begin{threeparttable}
\rowcolors{1}{}{gray!20}
\begin{tabularx}{0.6\textwidth}{llC} \toprule
Estimator & Test & $p$-value \\ \bottomrule \showrowcolors
Ordered probit & Joint Elo--TM slopes & 0.389 \\
Ordered probit & Elo slope & 0.337 \\
Ordered probit & Transfermarkt slope & 0.716 \\ \toprule
\end{tabularx}
\begin{tablenotes} \footnotesize
\item
\textit{Notes:}
The number of observations is 1394 in all cases, with 272 league phase matches.
The diagnostics test whether Elo and Transfermarkt slopes (interaction coefficients) differ in the 2024/25 league phase matches. Models include the league-phase indicator, its interactions with both strength indicators, and the baseline controls and fixed effects.
Significance: * $p<0.10$, ** $p<0.05$, *** $p<0.01$.
\end{tablenotes}
\end{threeparttable}
\end{subtable}
\end{table}

Table~\ref{Table8} tests whether the Elo and Transfermarkt slopes differ for the matches played in the 2024/25 league phase. None of the individual or joint interaction tests rejects the assumption of equal slope; even the smallest reported $p$-value exceeds 0.2.
However, the diagnostic is based on a single season and may have limited power to identify moderate changes. Hence, the result should be read as an absence of evidence for a break, not as proof that the relationship between the strength indicators and match outcomes is invariant to the format change.

\section{Conclusions} \label{Sec6}

Our paper has compared the predictive power of pre-season Elo ratings and lagged Transfermarkt values in recent seasons of UEFA club competitions. The two indicators achieve similar aggregate accuracy when used separately. Estimated stacking records the best RMSE for goal difference and expected goals difference, while equal-weight pooling has the best RPS under both probability models. However, these gains are small, and no combination dominates across all targets and evaluation seasons.
Equal-weight pooling retains the best RPS under every sample restriction. Its advantage in RPS over Elo is significant under the reported test, but the comparison with Transfermarkt remains uncertain.

According to \citet{Peeters2018}, Transfermarkt valuations robustly outperform Elo ratings for matches played by national teams. We find no clear Transfermarkt advantage in UEFA club competitions. The substantially higher number of club matches probably makes Elo ratings more informative, but our methodology is not able to not capture this effect. Differences in forecast horizon and the use of pre-season ratings and values also limit direct comparisons between the two studies.

To summarise, the objective, strictly performance-based Elo rating seems to be a sufficiently good predictor in club football. Even though combining this measure with market value could achieve a significant improvement in accuracy in certain settings, seeding is expected to remain close to optimal if it is based on Elo ratings, which is worth considering by tournament organisers.

\section*{Acknowledgements}
\addcontentsline{toc}{section}{Acknowledgements}
\noindent
We appreciate the help of \emph{Andr\'as Gyimesi} and the \emph{Social Science Computing Unit of the Databank at the ELTE Centre for Economic and Regional Studies} in data collection. \\
The research was supported by the National Research, Development and Innovation Office under Grants Advanced 152220 and FK 145838, and by the J\'anos Bolyai Research Scholarship of the Hungarian Academy of Sciences.

\bibliographystyle{apalike}
\bibliography{All_references}

\clearpage
\setcounter{section}{0}
\renewcommand{\thesubsection}{A.\arabic{subsection}}

\setcounter{figure}{0}
\renewcommand{\thefigure}{A.\arabic{figure}}

\setcounter{table}{0}
\renewcommand{\thetable}{A.\arabic{table}}

\section*{Appendix}
\addcontentsline{toc}{section}{Appendix}

\subsection{Calibration} \label{Sec_A1}

Table~\ref{Table_A4} presents binned calibration of home win probabilities for the combined ordered probit forecasts. For each specification, the observed frequency of home win exceeds the mean predicted probability in every bin. For the equal-weight pool, the gap is the smallest in the central probability range and larger in the 0.2--0.4 and 0.8--1.0 bins, but the latter contains only 31 matches.

\subsection{Additional specification checks} \label{Sec_A2}

For the Elo functional form check, we use the expected winning probability
\begin{equation}
W_{it}(h) = \frac{1}{1 + 10^{-(\text{Elo}_{it}+hH_{it})/400}},
\end{equation}
where \(h\) is the home advantage parameter (in Elo points) and $H_{it}$ is the home venue indicator. Neutral fixtures receive no adjustment as $H_{it}=0$. At each forecast origin, we select $\widehat h_t \in \{-50,-49,\ldots,150\}$ to minimise mean squared error against the home score (win $=1$, draw $=0.5$, loss $=0$) on earlier matches with unrestricted attendance in the common sample.

The raw Elo and $h=0$ specifications include separate \(H_{it}\) and \(H_{it}C_{it}\) terms. With estimated \(h\), the transformation contains the home adjustment, so we omit the separate \(H_{it}\) term and retain \(H_{it}C_{it}\). All three specifications include competition and phase fixed effects. The transformed score enters the outcome models as a predictor. According to Table~\ref{Table_A5}, the raw Elo difference performs slightly better for all three forecast targets.



Table~\ref{Table_A6} compares the baseline crowd-restriction flag with a season-wide classification of 2020/21 and 2021/22. Under the baseline definition, the entire 2020/21 season is treated as restriction-affected and eight verified closed-door matches are added from the 2021/22 season. The estimated restricted home effect remains positive under both definitions, but its magnitude and precision vary across outcomes. This exercise is therefore interpreted as a check on the home-context controls, not as a separate estimate of the effect of restrictions on attendance.

\begin{table}[t!]
\centering
\caption{Estimated Elo--Transfermarkt combination parameters}
\label{Table_A1}

\begin{subtable}{\textwidth}
\centering
\caption{Goal difference forecasts}

\begin{tabularx}{\textwidth}{l CcCCc} \toprule
Estimator & Forecast season & $N$ & Elo parameter & TM parameter & Intercept \\ \midrule
Stacked & 2023/24 &       238 &     0.608 &     0.466 &     0.234 \\
Stacked & 2024/25 &       476 &     0.587 &     0.454 &     0.268 \\ \bottomrule
\end{tabularx}
\end{subtable}

\vspace{0.5cm}
\begin{subtable}{\textwidth}
\centering
\caption{Expected goals difference forecasts}

\begin{tabularx}{\textwidth}{l CcCCc} \toprule
Estimator & Forecast season & $N$ & Elo parameter & TM parameter & Intercept \\ \midrule
Stacked & 2023/24 &       238 &     0.479 &     0.541 &     0.082 \\
Stacked & 2024/25 &       476 &     0.507 &     0.442 &     0.157 \\ \bottomrule
\end{tabularx}
\end{subtable}

\vspace{0.5cm}
\begin{subtable}{\textwidth}
\centering
\caption{Ordered match result forecasts}

\rowcolors{1}{}{gray!20}
\begin{threeparttable}
\begin{tabularx}{0.9\textwidth}{l ccCC} \toprule
Estimator & Forecast season & $N$ & Elo parameter & TM parameter \\ \bottomrule
Ordered probit & 2023/24 &       238 &     0.6 &     0.4 \\
Ordered probit & 2024/25 &       476 &     0.65 &     0.35 \\ \hline
Multinomial logit & 2023/24 &       238 &     0.65 &     0.35 \\
Multinomial logit & 2024/25 &       476 &     0.65 &     0.35 \\ \bottomrule
\end{tabularx}
\begin{tablenotes} \footnotesize
\item \textit{Notes:} For continuous outcomes, entries are unconstrained stacking coefficients estimated from earlier out-of-sample forecasts and need not sum to one. Column Intercept reports the constant in each stacking regression; probability pools have no intercept. For the ordered match result probabilities, the parameters are convex Elo--Transfermarkt mixture weights selected by minimising previous out-of-sample RPS over a grid.
\end{tablenotes}
\end{threeparttable}
\end{subtable}

\end{table}

\begin{table}[t!]
\centering
\caption{Forecast performance by evaluation season}
\label{Table_A2}

\begin{subtable}{\textwidth}
\centering
\caption{One season forecast: 2023/24}

\rowcolors{1}{gray!20}{}
\begin{threeparttable}
\begin{tabularx}{\textwidth}{l CCCC} \toprule \hiderowcolors
Model & Goal difference & $\mathit{xG}$ difference & Ordered probit & Multinomial logit \\ 
Measure of loss & RMSE & RMSE & $100 \times$RPS & $100 \times$RPS \\ \bottomrule \showrowcolors
Historical &     1.830 &     1.329 &    22.997 &    22.997 \\
Elo &     1.531 &     1.141 &    18.287 &    18.318 \\
Transfermarkt &     1.537 &     1.140 &    18.509 &    18.518 \\
Joint Elo--TM &     1.522 &     1.134 &    18.185 &    18.212 \\
Equal-weight pool &     1.519 &     1.129 &    18.246 &    18.273 \\
Estimated combination &     1.489 &     1.118 &    18.230 &    18.256 \\ \bottomrule
\end{tabularx}
\begin{tablenotes} \footnotesize
\item \textit{Notes:}
The number of observations is 238 in all cases. Lower values are better.
\end{tablenotes}
\end{threeparttable}
\end{subtable}

\vspace{0.5cm}
\begin{subtable}{\textwidth}
\centering
\caption{One season forecast: 2024/25}

\rowcolors{1}{gray!20}{}
\begin{threeparttable}
\begin{tabularx}{\textwidth}{l CCCC} \toprule \hiderowcolors
Model & Goal difference & $\mathit{xG}$ difference & Ordered probit & Multinomial logit \\ 
Measure of loss & RMSE & RMSE & $100 \times$RPS & $100 \times$RPS \\ \bottomrule \showrowcolors
Historical &     1.830 &     1.329 &    22.997 &    22.997 \\
Elo &     1.531 &     1.141 &    18.287 &    18.318 \\
Transfermarkt &     1.537 &     1.140 &    18.509 &    18.518 \\
Joint Elo--TM &     1.522 &     1.134 &    18.185 &    18.212 \\
Equal-weight pool &     1.519 &     1.129 &    18.246 &    18.273 \\
Estimated combination &     1.489 &     1.118 &    18.230 &    18.256 \\\bottomrule
\end{tabularx}
\begin{tablenotes} \footnotesize
\item \textit{Notes:}
The number of observations is 238 in all cases. Lower values are better.
\end{tablenotes}
\end{threeparttable}
\end{subtable}

\end{table}

\begin{table}[t!]
\caption{Selected pairwise comparison tests of forecasts}
\label{Table_A3}

\begin{subtable}{\textwidth}
\caption{Goal difference, squared error loss}

\rowcolors{1}{}{gray!20}
\begin{tabularx}{\textwidth}{lCCcC} \toprule
Comparison & $\Delta$ & Std.e. & 95\% CI & $p$-value \\ \bottomrule
Equal-weight vs.\ Elo & \(-0.052\) &     0.032 & \([-0.115,\,+0.010]\) & 0.099* \\
Equal-weight vs.\ TM & \(-0.052\) &     0.032 & \([-0.116,\,+0.011]\) & 0.105 \\
Estimated vs.\ Equal-weight & \(-0.031\) &     0.039 & \([-0.107,\,+0.046]\) & 0.431 \\ \toprule
\end{tabularx}
\end{subtable}

\vspace{0.5cm}
\begin{subtable}{\textwidth}
\caption{Expected goals difference, squared error loss}

\rowcolors{1}{}{gray!20}
\begin{tabularx}{\textwidth}{lCCcC} \toprule
Comparison & $\Delta$ & Std.e. & 95\% CI & $p$-value \\ \bottomrule
Equal-weight vs.\ Elo & \(-0.026\) &     0.016 & \([-0.058,\,+0.006]\) & 0.116 \\
Equal-weight vs.\ TM & \(-0.034\) &     0.017 & \([-0.067,\,-0.001]\) & 0.046** \\
Estimated vs.\ Equal-weight & \(-0.009\) &     0.012 & \([-0.032,\,+0.015]\) & 0.476 \\ \toprule
\end{tabularx}
\end{subtable}

\vspace{0.5cm}
\begin{subtable}{\textwidth}
\caption{Ordered probit, log loss}

\rowcolors{1}{}{gray!20}
\begin{tabularx}{\textwidth}{lCCcC} \toprule
Comparison & $\Delta$ & Std.e. & 95\% CI & $p$-value \\ \bottomrule
Equal-weight vs.\ Elo & \(-0.009\) &     0.005 & \([-0.018,\,+0.001]\) & 0.080* \\
Equal-weight vs.\ TM & \(-0.004\) &     0.005 & \([-0.013,\,+0.005]\) & 0.375 \\
Estimated vs.\ Equal-weight & \(+0.001\) &     0.001 & \([-0.001,\,+0.004]\) & 0.316 \\ \toprule
\end{tabularx}
\end{subtable}

\vspace{0.5cm}
\begin{subtable}{\textwidth}
\caption{Multinomial logit, RPS loss}

\begin{threeparttable}
\rowcolors{1}{}{gray!20}
\begin{tabularx}{\textwidth}{lCCcC} \toprule
Comparison & $100 \times \Delta$ & $100\times$Std.e. & 95\% CI & $p$-value \\ \bottomrule
Equal-weight vs.\ Elo & \(-0.309\) &     0.146 & \([-0.596,\,-0.021]\) & 0.035** \\
Equal-weight vs.\ TM & \(-0.072\) &     0.148 & \([-0.364,\,+0.220]\) & 0.627 \\
Estimated vs.\ Equal-weight & \(+0.053\) &     0.044 & \([-0.034,\,+0.139]\) & 0.232 \\ \toprule
\end{tabularx}
\begin{tablenotes} \footnotesize
\item
\textit{Notes:}
The number of observations is 600 in all cases.
$\Delta$ is the score of model A minus the score of model B; negative values favour model A. Standard errors are clustered by unordered team-pair-season within competition. The 95\% confidence intervals use the same $t$ distribution and degrees of freedom as the reported tests. Equal-weight and Estimated denote uniform and estimated forecast pools. The intervals condition on the fitted forecasts; they do not re-estimate models or pooling weights.
Significance: * $p<0.10$, ** $p<0.05$, *** $p<0.01$.
\end{tablenotes}
\end{threeparttable}
\end{subtable}

\end{table}

\begin{table}[t!]
\centering
\caption{Home win probability calibration, main sample}
\label{Table_A4}

\begin{subtable}{\textwidth}
\caption{Joint Elo--Transfermarkt}
\centering

\rowcolors{1}{}{gray!20}
\begin{tabularx}{0.8\textwidth}{lcCC} \toprule
Probability bin & $N$ & Mean predicted & Observed frequency \\ \bottomrule
\([0.0,\,0.2)\) &        72 &     0.133 &     0.236 \\
\([0.2,\,0.4)\) &       167 &     0.306 &     0.365 \\
\([0.4,\,0.6)\) &       188 &     0.493 &     0.521 \\
\([0.6,\,0.8)\) &       131 &     0.700 &     0.710 \\
\([0.8,\,1.0]\) &        42 &     0.864 &     0.881 \\ \toprule
\end{tabularx}
\end{subtable}

\vspace{0.5cm}
\begin{subtable}{\textwidth}
\caption{Equal-weight pool}
\centering

\rowcolors{1}{}{gray!20}
\begin{tabularx}{0.8\textwidth}{lcCC} \toprule
Probability bin & $N$ & Mean predicted & Observed frequency \\ \bottomrule
\([0.0,\,0.2)\) &        55 &     0.139 &     0.200 \\
\([0.2,\,0.4)\) &       172 &     0.306 &     0.384 \\
\([0.4,\,0.6)\) &       210 &     0.489 &     0.500 \\
\([0.6,\,0.8)\) &       132 &     0.695 &     0.720 \\
\([0.8,\,1.0]\) &        31 &     0.856 &     0.935 \\ \toprule
\end{tabularx}
\end{subtable}

\vspace{0.5cm}
\begin{subtable}{\textwidth}
\caption{Joint Elo--Transfermarkt}
\centering

\begin{threeparttable}
\rowcolors{1}{}{gray!20}
\begin{tabularx}{0.8\textwidth}{lcCC} \toprule
Probability bin & $N$ & Mean predicted & Observed frequency \\ \bottomrule
\([0.0,\,0.2)\) &        55 &     0.134 &     0.182 \\
\([0.2,\,0.4)\) &       175 &     0.303 &     0.377 \\
\([0.4,\,0.6)\) &       205 &     0.491 &     0.507 \\
\([0.6,\,0.8)\) &       134 &     0.698 &     0.731 \\
\([0.8,\,1.0]\) &        31 &     0.861 &     0.903 \\ \toprule
\end{tabularx}
\begin{tablenotes} \footnotesize
\item \textit{Notes:}
Bins are based on predicted home win probabilities by the ordered probit model.
\end{tablenotes}
\end{threeparttable}
\end{subtable}

\end{table}

\begin{table}[t!]
\centering

\caption{Robustness to Elo functional form}
\label{Table_A5}

\begin{subtable}{\textwidth}
\centering
\caption{Out-of-sample forecast performance}

\begin{threeparttable}
\rowcolors{1}{gray!20}{}
\begin{tabularx}{\textwidth}{lCCC} \toprule \hiderowcolors
Target & Goal difference & $\mathit{xG}$ difference & Ordered result \\
Measure of loss & RMSE & RMSE & $100 \times$RPS \\ \bottomrule \showrowcolors
Raw Elo difference &     1.710 &     1.183 &    19.729 \\
Elo expected prob., $h=0$ &     1.714 &     1.189 &    19.806 \\
Elo expected prob., estimated $h$ &     1.721 &     1.193 &    19.850 \\ \toprule
\end{tabularx}
\begin{tablenotes} \footnotesize
\item
\textit{Notes:}
The number of observation is 600. RPS is multiplied by 100.
\end{tablenotes}
\end{threeparttable}
\end{subtable}

\vspace{0.5cm}
\begin{subtable}{\textwidth}
\centering
\caption{Home advantage parameter estimated from prior seasons}

\begin{threeparttable}
\rowcolors{1}{}{gray!20}
\begin{tabularx}{0.8\textwidth}{lCCc} \toprule
Forecast season & Calibration $N$ & Estimated $h$ & Calibration MSE \\ \bottomrule
2022/23 &       248 &      40 &   0.15041 \\
2023/24 &       486 &      66 &   0.14897 \\
2024/25 &       724 &      73 &   0.14668 \\ \toprule
\end{tabularx}
\begin{tablenotes} \footnotesize
\item
\textit{Notes:}
The home adjustment parameter $h$ in Elo points is estimated at each forecast origin using only earlier unrestricted-attendance matches.
\end{tablenotes}
\end{threeparttable}
\end{subtable}
\end{table}


\begin{table}[t!]
\centering
\caption{Home advantage under alternative crowd-restriction definitions}
\label{Table_A6}

\begin{subtable}{\textwidth}
\caption{Goal difference}

\begin{tabularx}{\textwidth}{lCCCc} \toprule
Restriction definition & Unrestricted & Adjustment & Restricted & $p$-value \\ \midrule
Baseline crowd-restriction &     0.416 &     0.281 &     0.697 &     0.052 \\
2020/21--2021/22 seasons &     0.488 &     0.064 &     0.552 &     0.119 \\ \bottomrule
\end{tabularx}
\end{subtable}

\vspace{0.5cm}
\begin{subtable}{\textwidth}
\caption{Expected goals difference}

\begin{tabularx}{\textwidth}{lCCCc} \toprule
Restriction definition & Unrestricted & Adjustment & Restricted & $p$-value \\ \midrule
Baseline crowd-restriction &     0.508 &    $-$0.070 &     0.438 &     0.099 \\
2020/21--2021/22 seasons &     0.716 &    $-$0.587 &     0.129 &     0.619 \\ \bottomrule
\end{tabularx}
\end{subtable}

\vspace{0.5cm}
\begin{subtable}{\textwidth}
\caption{Ordered result latent index}

\begin{threeparttable}
\begin{tabularx}{\textwidth}{lCCCc} \toprule
Restriction definition & Unrestricted & Adjustment & Restricted & $p$-value \\ \midrule
Baseline crowd-restriction &     0.071 &     0.314 &     0.385 &     0.194 \\
2020/21--2021/22 seasons &     0.090 &     0.219 &     0.309 &     0.327 \\ \bottomrule
\end{tabularx}
\begin{tablenotes} \footnotesize
\item \textit{Notes:}
The number of observations is 1394 in all cases.
Column Unrestricted is the coefficient of the home venue indicator. Column Adjustment is its interaction with the stated crowd-restriction definition. Column Restricted is their sum. OLS coefficients are for goals or $\mathit{xG}$, while the ordered probit coefficient is for the latent outcome index. The $p$-values refer to tests whether the restricted home effect equals zero.
\end{tablenotes}
\end{threeparttable}
\end{subtable}

\end{table}

\end{document}